\documentclass[runningheads]{llncs}
\usepackage{graphicx}
\begin{document}
\title{A Comparative Study of Younger and Older Adults' Interaction with a Crowdsourcing Android TV App for Detecting Errors in TEDx Video Subtitles}
%\thanks{Supported by organization x.}
%
\titlerunning{Interaction with an App for Detecting Errors in Subtitles}
% If the paper title is too long for the running head, you can set
% an abbreviated paper title here
%
\author{Kinga Skorupska\inst{1,2}\orcidID{0000-0002-9005-0348} \and
Manuel N\'{u}\~{n}ez\inst{1}\orcidID{0000-0002-1197-7488} \and
Wies\l{}aw Kope\'{c}\inst{1}\orcidID{0000-0001-9132-4171} \and
Rados\l{}aw Nielek\inst{1}\orcidID{0000-0002-5794-7532}}

\authorrunning{Skorupska et al.}
% First names are abbreviated in the running head.
% If there are more than two authors, 'et al.' is used.
%
\institute{Polish-Japanese Academy of Information Technology \\
\email{kinga.skorupska@pja.edu.pl}\\
\url{http://www.pja.edu.pl}}
\maketitle              % typeset the header of the contribution
\begin{abstract}
In this paper we report the results of a pilot study comparing the older and younger adults' interaction with an Android TV application which enables users to detect errors in video subtitles. Overall, the interaction with the TV-mediated crowdsourcing system relying on language proficiency was seen as intuitive, fun and accessible, but also cognitively demanding; more so for younger adults who focused on the task of detecting errors, than for older adults who concentrated more on the meaning and edutainment aspect of the videos. We also discuss participants' motivations and preliminary recommendations for the design of TV-enabled crowdsourcing tasks and subtitle QA systems.

\keywords{Crowdsourcing  \and Smart TV \and Android TV\and Design evaluation\and Subtitles \and Older adults \and Younger adults.}
\end{abstract}

\section{Introduction and related works}

With the increasing amount of video content it is necessary to ensure its accessibility to the deaf, the hard of hearing and international audiences through quality same language and multilingual subtitles. Therefore, crowdsourcing subtitle quality assurance (QA) models are an important research frontier, especially as subtitles are often created by volunteers, as in the case of TED and TEDx \cite{camara2014multilingual} or generated automatically. At the same time, there are groups who may benefit from more fun and accessible crowdsourcing projects. 

For example, older adults, who comprised 19.2\% of the EU-28 population in 2016 \cite{population_structure2017}, % of ....
% stats
benefit from all forms of volunteering, as it slows the negative effects of aging and helps combat depression \cite{lum2005effects}. Yet, there exist multiple barriers to their inclusion in typical crowdsourcing tasks, such as lower ICT skills, uncomfortable and costly setup of such solutions \cite{sandhu2013ict}, unfamiliar interfaces and lack of motivation due to unclear personal benefit \cite{brewer2016would}, unsocial nature of the task \cite{vines2011eighty} or their perception of not being qualified \cite{kopec2017older}. 

Younger adults, on the other hand, who are more open to online crowdsourcing and microtasking, comprise a significant number of online video viewers, as, according to We Are Flint about 96\% of people in UK and US aged 18-34 watch YouTube videos \cite{Youtubestats}. Both groups are relevant to the development of TV-enabled subtitle QA crowdsourcing tasks as potential contributors and audience.

Therefore, the key research goal was to validate a novel interface for creating no-grind crowdsourcing solutions, ones that do not rely on tedious repetition, with two relevant user groups. To do this, we deployed a Smart TV-based system based on best practices of designing for older users \cite{fiskdesigningolderr} \cite{pak_mclaughlin_2011} with a comfortable at-home setup, large screen size, and remote relying on familiar interaction patterns \cite{pan2015effects} with engaging edutainment crowdsourcing tasks. This lowered ICT and other participation barriers and allowed us to signal some possible differences in the participants' approach, motivation, mode of use, experience and expectations. We lay ground to the discussion of the extent to which one may build a universal crowdsourcing system suited to the needs of these different groups, to tap into their potential, facilitate social inclusion and build social capital. 

\section{Methods}

\subsection{Comparative study design}

To explore these considerations we conducted a comparative qualitative study in the course of which we compared results from a study involving older adults \cite{Skorupska:2018:OAC:3290265.3274428} to the results of a study with younger adults conducted in February-March 2019.

\begin{figure}
    \includegraphics[width=340px]{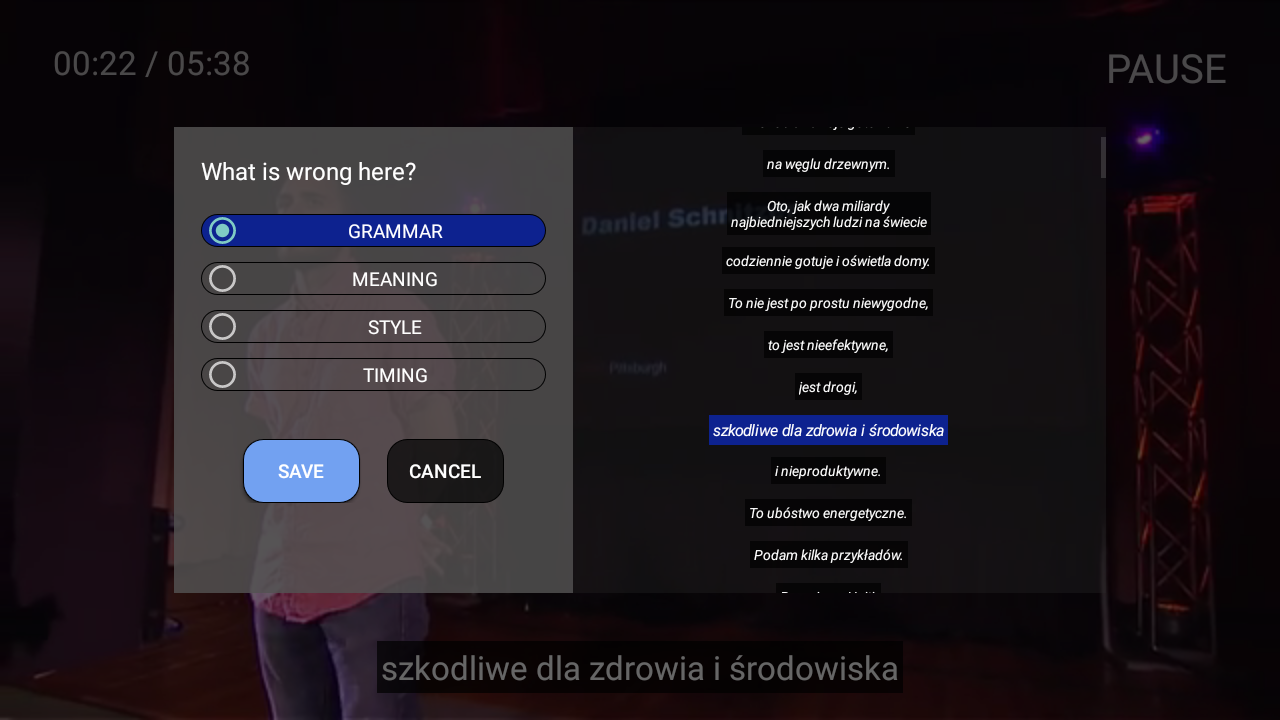}
    \caption{The error category selection overlay in our Dream TV application}
    \label{fig:interface}
\end{figure}

The study examined the interaction with the DreamTV application we created \cite{Skorupska:2018:OAC:3290265.3274428} which allows users to watch TEDx videos with volunteer-created subtitles retrieved from Amara API. Once they spot an error they can pause the video to display an overlay (Fig. \ref{fig:interface}) where they choose the error category among grammar, meaning, style and timing. These error categories were chosen based on preliminary tests and research to be more intuitive than existing models of quality assessment of subtitles by professionals \cite{romero2018quality} and to aid in improving the subtitles later within the pipeline or during post-editing.

The research protocol, which took about two hours to complete, involved individual testing at participants' homes, where an Android TV set-top box was connected to participants' TV sets, to provide the most natural use conditions, as proposed in multiple studies on Living Labs \cite{kopec2017living} \cite{alaoui2013livinglab}. It consisted of the DigComp survey\footnote{A survey measuring indicators of Digital Competence based on the Digital Competence Framework \cite{ferrari2013digcomp}.}, a semi-structured interview to evaluate experience with subtitles, the explanation of the project, that is the study and its benefits, an introduction to subtitles and a subtitle error detection written exercise, an app demonstration and a hands-on test, free interaction with the application and our pre-selected test videos (two in Polish, three in English) with redacted Polish subtitles.

For our study we selected five videos to represent different challenges. They were controlled for topic, length, source language (spoken), ease of comprehension and errors: saturation, category and source, either machine (using SubtitleEdit and Google Translate) or organic human or introduced by researchers based on common errors lists on TED Translators' wiki\footnote{The TED Translators' wiki containing lists of common errors can be found at: https://translations.ted.com}.
The videos selected and errors introduced allowed us to observe a variety of factors at play, in order to gather diverse insights to determine interesting areas of further inquiry.

\subsection{Participants}

We invited seven older adults (O1-O7) and seven younger adults (Y1-Y7) to participate in our study, in each case three female participants and four male participants. We controlled for age, occupation and ICT skills ("above basic proficiency", which is the highest level in DigComp). All participants live in Warsaw, the capital city of Poland. For older adults all owned TVs, including two Smart TVs, and had a dedicated entertainment space in their living room. There was a 20 years age span: the youngest participant was 60 years old and the oldest one was 79, mean 70.85 (SD=6.87). For younger adults we recruited a group that would share the most relevant characteristics with our older adults, especially in terms of their housing situation and entertainment setup, which meant that in Poland they had to be between 25-35 years of age. All but one participants owned Smart TVs and had their own dedicated entertainment space in the living room. All were professionally active and none of them had children. The age span was 5 years, as the youngest participant was 28, and the oldest 33, mean 30.71 (SD=2.28).

\section{Results and discussion}

Overall, using the application was enjoyable, intuitive and easy for both younger and older participants, however there were differences in their approach to the task. While our group of younger adults saw it as an enjoyable activity one could do to improve subtitles, brag or supplement their income in a fun way, our group of older adults viewed it less as work and more an opportunity to learn something and did not expect payment for contributing. For older adults it was more interesting, as they were given access to resources they were unlikely to reach to on their own (TEDx videos) whereas younger adults agreed that they know less demanding or better entertainment. Younger adults detected more mistakes than older adults as they viewed the task to be more work-like and in consequence, demanding. Older adults seemed more lenient, especially when it came to style and punctuation, and focused more on the content of the videos, rather than correcting mistakes. There were also differences in feedback. Where older adults focused on ways to find videos that would be a better fit for them thematically, younger adults focused more on critiquing the error categories chosen and comparing the application to Netflix. This is due to the differences in experience with such services. Both groups found the interaction via the remote to be very convenient and well-suited for this activity and they learned to comfortably use the application in just one session, with older adults in general taking more time to learn and later to navigate, but with no significant other differences.

\subsection{Error detection}

\subsubsection{Reflexes} Overall, all of the older participants paused the videos one subtitle too late, and had to use the dialog list to navigate back to the subtitle where they wanted to mark the error. The same was true of all but one younger adults, as Y2 paused even before the speaker finished the sentence, indicating that they read rather than listened. This suggests that access to the full dialog list is necessary in this type of crowdsourcing for all age groups.

\subsubsection{Number of errors found} In general, younger adults found more errors than older adults which may be related to their attitude towards this activity. While younger adults focused on the task of finding errors, older adults engaged with the content of the videos more and felt that they are learning new interesting things (O1-O4). This is in contrast with younger adults, except for Y4, who admitted to focus more on the content and commented that they "should watch such videos more often as they are interesting". Consequently, younger adults found many more punctuation errors, which older adults often ignored. This may be as punctuation errors do not interfere with understanding. Older adults, who focused more on understanding the content, often chose the "meaning" category, when something was not clear to them (e.g. "it is not explained what is this photon" or "Spiderman, this is not Polish" by O2 and "kryptonite, must be a mistake" by O5, O6), suggesting the application could benefit from a built-in dictionary. Older adults' focus on meaning is in line with Radvansky's research on the effect of aging on memory and comprehension, suggesting that while lower levels of memory, which may be responsible for remembering specifics such as punctuation, deteriorate with age, the ability to form situation models on a higher level, aiding in meaning and general comprehension is less affected \cite{Radvansky99memorycomprehension}. Moreover, different people found very different errors, depending on their interests and background (science for Y6: "the Sun vs the sun", detailed punctuation rules for Y2 with linguistic background) which shows that the effect of scale by relying more on quantity and not quality of contributions may work well here.

\subsubsection{Error categories} All but one of the younger participants (Y1-Y6) encountered errors in subtitles to which they wished to assign more than one error category, to remove the analysis paralysis of choosing the best fitting category ("People like me would deliberate 3 years over a single word" Y1) and likely to satisfy their need for cognitive closure \cite{websterkrugalski95cognclosure}, as many younger participants found the categories to be "fuzzy". The other participant, Y7, said that "these are short lines so if someone marks a mistake it is easy to know what it is" and proposed to remove categories, the same could be seen in O3's eagerness to just mark mistakes quickly and continue watching the videos.

Younger adults remarked that "synchronization is the most intuitive" (Y1). Other error categories requested were "punctuation" (Y6) and "subtitle division" (line breaking) (Y3) and "technical errors" such as subtitle convention errors as a separate category (Y1, Y2) and both Y7 and Y3 said that knowing subtitle conventions requires a lot of practice, and pre-teaching, for which Y3 suggested a mini-game, while older adults wished for an in-application tutorial to ensure they do not make mistakes when marking mistakes (O1, O4). One participant, Y6, also said there ought to be a way to mark recurring errors ("Here I would have to mark a lot of things, because the Sun should be written with capital letter, and it repeats a lot"), on the other hand O3 remarked "He made the same mistake, but I'll overlook it now", eager to continue watching.

Older adults (O1-O7) did not question the error categories even though they often could not decide which category to choose (O4, O5) and sometimes deliberated aloud (O3). This may be because older adults are less likely to criticize design choices in the context of technology, as they feel they lack experience in it so they are not confident enough to know they can contribute. This was also observed in the context of participatory design by Kopec et al \cite{kopec2017older}. Also, even though some older adults had to sit closer to the screen to read (O1, O3) it was a younger adult (Y6) who voiced that they would like the interface to be bigger.

In conclusion, to ease the choice of error categories we propose to present them in the order of importance, with the top category being "meaning" - answering the question "Is this subtitle understandable?", followed by "grammar", as it includes common punctuation mistakes, and then "style", which would have to be explained as relating to technical errors, and including also other problems. We postulate that because of conflicts of simultaneous work it is very difficult to find synchronization errors, while also looking for other types of errors ("It is difficult to catch problems with synchronization - you focus on all the other mistakes" Y3, and "I had to read" O5). This was seen in the tests with older adults, who found no synchronization errors (O1-O7), and younger adults who rarely marked them as they found it tiring to both read, and listen (Y7: "I did not listen to the guy", Y6: "difficult to focus on what the person was saying") Signalling the relationship between enjoyment, interest and errors found Y6 said: "this topic was interesting, sometimes I did not focus on finding mistakes". Both older (O1-O4, O6) and younger adults (Y3, Y4, Y6) seemed to find fewer errors the more they enjoyed the video, with Y4 saying that they were "forgetting to read". The enjoyment was also negatively correlated with the number of errors marked, with Y2 saying that "The errors were so thickly distributed, it is a very tiring video" and that "If there were fewer errors it would be more fun than work" and Y5 mentioning that "If you have to focus only on subtitles it is more like work, but if you get to mark glaring errors only it is more entertainment".

\subsection{Fun or work?}

Y1 and Y5 found the application to be very fun, commenting that "you can point out someone's mistakes without arguing with that person, everyone loves that!" (Y1), adding that it is true especially when there are people around, and "How fun! I like it! I could do it all my life" (Y5). Y6 also said "it's cool, I like nitpicking". The other participants commented that it would be work if you "had to do it, like an editor in a paper" and "the movies are not long, and you can take breaks" (Y7). Similarly, Y3 mentioned that "you should be able to choose how long video you want". This aspect of controlling time was also present in older adults' feedback, as they enjoyed the ability to pause the video at will, take breaks, and O3 even said "The movies should be shorter, then I could watch anything! Just give me ten 5 minute films and I can do that for an hour". Older adults overall focused on the educational aspect of the task, saying that it is good practice and one can "learn a lot" (O1-O4) from these videos. This aspect was less prominent with younger adults, who often treated the experience almost job-like as it was "mentally demanding" and felt more like "work", or that it is a bit like an "exam" (Y1) and felt judged when they did not understand a subtitle (Y3) ("I don't know what they mean by "last mile" and since it was in quotation marks it must be something that everyone knows, so now I feel stupid"). In contrast, only O4 mentioned that "It is tiring, I am not that young anymore." drawing attention to the task's cognitive load.

\subsection{Motivation, Gamification and Rewards}

\begin{table}
 \caption{Comparison of older and younger adults' motivations, rewards and wishes}
  \label{tab:table1}
\begin{tabular}{p{3.5cm}|p{4.5cm}|p{4cm}}
%\hline
\textbf{} & \textbf{Younger adults} & \textbf{Older adults}\\
\hline
\textbf{Pointing out mistakes} & Y1, Y2, Y5, Y6 & O3\\
  \hline
\textbf{Social activity} & Y1: "to do with friends" &  O2: "with grandchildren" \\   \hline
\textbf{Helping somebody} & Y1: "If some friend asked me to do this for them, I would help them", Y4 &  \\
  \hline
\textbf{Learning new things} & Our group of younger adults could watch such videos, but just watch as Y4: "they are interesting" to Y6: "focus on the content". & O1-O4, O1: "I learned a lot", O3 "I would watch movies about health, global issues, climate change or politics" but: O5 "The topics would have to be useful"\\
  \hline
\textbf{Getting paid} & Y1-Y7, except for Y5: "Nobody would pay much, it's better to have bonuses, like a subscription or a small gift because earning little money is meh" &  \\
  \hline
\textbf{Improving the world} & Y1: "I like it, if I was convinced myself that this is making the world a bit better, then this is a convenient way to help" &   \\
  \hline
\textbf{Challenging oneself cognitively} & Y2 and Y3, but about other people, Y3: "blue-collar workers" and "stay at home moms" who can do it for fun and Y2: "retired people to stay active". & O3: "This task is great for old people, but only those who are mentally fit, so that they don’t deteriorate" \\
\hline
\textbf{Passion for the topic} & Y3 mentioning feminists: "people who are very passionate about a topic can contribute" & 
\\
  \hline
\textbf{Statistics of confirmed contributions} & Y4: "ranking like on Memrise", Y6: "ranking of best reviewers", Y3: "a community to care about my achievements listed on my profile". Interestingly, both Y1 and Y2 mentioned they do not need statistics. &  \\ \hline
\textbf{Helping improve subtitles being used} & Y3: "that there were 100 people who watched this film with improved subtitle in a month would mean something" &  \\   \hline
\textbf{Access to training} & Y3, "in the community access to games that help you develop  skills to contribute better" & O1: "It would be good to have a testing mode, to be able to train without consequences", O4\\   \hline
\textbf{Addressing glaring errors }& in videos they are watching with subtitles anyway (Y4, Y5) &   \\\hline
\textbf{Reliance on linguistic experience} &  Y3: "I like that I don't have to learn anything to start doing it, I know the language" & O3: "There should be more subtitle testers like me, but not young people because they have little experience"  \\ 
\hline
\end{tabular}
\end{table}

While older adults' participants motivation was mostly based on the value for them, in terms usefulness, relevance to their interests and staying active, for younger adults there was almost no concern about the topic as they viewed the task to be more "work-like" and focused finding errors more than understanding and enjoying the content - likely because they have other entertainment readily available. Detailed comparison of approaches and attitudes is visible in Table \ref{tab:table1}.

\subsection{Sustainability}

Overall, although most of the participants found this activity to be fun, there are doubts whether they would do it in the long run without other incentives. The tests with older adults suggest that some may continue using the application as an easy foray into the world of edutainment and to stay active, except for O5 who stated "I manage, but it is not my thing - the topics would have to be useful" and O4 who expected to be bored as one has to "be focused". On the other hand, some younger adults commented "I wouldn't do it because it is time consuming, when you watch something to gain knowledge it is easier to understand the content if you are just watching" (Y6) or "it's not my type of thing, I am not a linguist and correcting errors is not my passion" (Y7). They also mentioned shortage of time (Y2, Y3) and the demanding nature of this task (Y2, Y3, Y6, Y7) as a problem.
For younger adults, who have formed habits regarding their access to other forms of entertainment, it may work best as a feature integrated into their familiar experience. Both Y4 and Y5 suggested that such activity could be "integrated into a player" they use anyway", on YouTube for Y4 ("it could be great if YouTube had something like that in their automatic subtitles, which now suck") or on VOD for Y5, who noted that "Sometimes I am tempted to mark something on VOD - there are few people who would bother to go to a film distributors' website and report errors in subtitles". Y5 concluded that "If it was easily accessible then a lot of people would do it, if they could just mark something on their remote".

\section{Conclusions}

As this is a pilot study with a small number of participants it is important to verify the following preliminary findings. While this task is fun for both younger and older adults, the former treat it more like work and expect payment. This group would benefit from having a similar solution integrated into their entertainment medium of choice. On the other hand, older adults are a promising target for this type of crowdsourcing, as it not only provides them with content they may otherwise miss, but also allows them to learn and stay active.

Future work ought to explore TV-mediated crowdsourcing in larger studies, and focus on the patterns of interaction with this solution, including the timing of engagement and quantitative relationship between the enjoyment of the video and the number of subtitle errors found. It is also important to verify if this TV-mediated crowdsourcing solution can hold older adults' interest over time, and if so, what are other ways such mode of interaction can be used to allow older adults to stay active for longer, contribute to society and learn new things.

\textit{This research in part was supported by the Polish National Science Center grant 2018/29/B/HS6/02604 and the European Union’s Horizon 2020 research and innovation programme under the Marie Skłodowska-Curie grant agreement No 690962.}

\bibliographystyle{splncs04}
\bibliography{bibdream.bib}

\end{document}